\documentclass{PoS}
\usepackage[iso-8859-7]{inputenc}
\usepackage{epsfig}
\usepackage{graphicx}
\usepackage{epstopdf}
\usepackage{amsfonts}
\usepackage{amssymb}
\usepackage{amsthm}
\usepackage{amsmath}
\usepackage{multirow}
\newcommand{\newc}{\newcommand}
\newc{\ra}{\rightarrow}
\newc{\lra}{\leftrightarrow}
\newc{\be}{\begin{equation}}
\newc{\ee}{\end{equation}}
\newc{\bs}{\begin{split}}
	\newc{\es}{\end{split}}
\newc{\ba}{\begin{eqnarray}}
\newc{\ea}{\end{eqnarray}}
\newc{\ov}{\overline}
\newc{\pa}{\partial}
\newc{\D}{\Delta}

\newc{\nn}{\nonumber}

\title{Moduli Stabilization and Inflation in Type IIB/F-theory}

\ShortTitle{F-inflation}

       \author{Ignatios Antoniadis\\
            Laboratoire de Physique Th\'eorique et Hautes \'Energies - LPTHE,\\
            	Sorbonne Universit\'e, CNRS, 4 Place Jussieu, 75005 Paris, France\\
            	and\\
              	Albert Einstein Center, Institute for Theoretical Physics, University of Bern,
             \\
               			Sidlerstrasse 5, CH-3012 Bern, Switzerland\\
               			           E-mail: \email{antoniad@lpthe.jussieu.fr}}

\author{Yifan Chen\\
          Laboratoire de Physique Th\'eorique et Hautes \'Energies - LPTHE,\\
                  	Sorbonne Universit\'e, CNRS, 4 Place Jussieu, 75005 Paris, France\\
                    E-mail: \email{yifan.chen@lpthe.jussieu.fr}}

              \author{\speaker{George K. Leontaris}\\
                  		Physics Department, University of Ioannina\\
                   		 45110, Ioannina, 	Greece\\
                      E-mail: \email{leonta@uoi.gr}}

\abstract{In the first part of this talk, a short overview of the ongoing debate on the existence of de Sitter vacua in string theory 
is presented. In the second part, the moduli stabilisation and inflation are discussed in the context of type IIB/F-theory. Considering a 
 configuration of three intersecting $D7$ branes with fluxes, it is shown that   higher loop effects inducing logarithmic corrections 
 to the K\"ahler potential can stabilise the K\"ahler moduli in a de Sitter Vacuum. When a new Fayet-Iliopoulos term is included,  it is   also possible 
 to generate the required number of e-foldings and satisfy the conditions for  slow-roll inflation. }

\FullConference{Corfu Summer Institute 2018 "School and Workshops on Elementary Particle Physics and Gravity"\\
		(CORFU2018)\\
		31 August - 28 September, 2018\\
		Corfu, Greece}

\begin{document}

\section{Introduction}

Twenty years ago a big breakthrough has been made in cosmology as a result of the significant 
observational discovery ~\cite{Riess:1998cb,Perlmutter:1998np}, indicating that there is an ongoing accelerating expansion 
of the universe which started before its thermalisation. According to the standard interpretation
this implies that the universe is entering an era dominated  with  dark energy permeating  all of space.  In the framework of 
Einstein's general relativity, dark energy can  be accounted for by a positive value of the cosmological constant $\Lambda$ 
with a present day value of $\Lambda \approx 10^{-120} M_{P}^4$ where $M_{P}$ is the four-dimensional Planck mass.
 In the context of effective field theories, the simplest scenario describing  the essential  features of these  facts, 
 consists of a scalar field $\phi$ acquiring a potential $V(\phi)$  with  positive vacuum energy equal to the cosmological constant. 
That is, a model with a scalar potential possessing a stable or metastable de Sitter (dS) vacuum.  
Provided some additional  requirements are fulfilled, the potential $V(\phi)$ could be appropriate to  successfully realise slow 
roll inflation with the scalar $\phi$ field playing the r\^ole of the inflaton field. 
 \begin{figure}[h!]
	\centering
	\includegraphics[width=0.50\columnwidth]{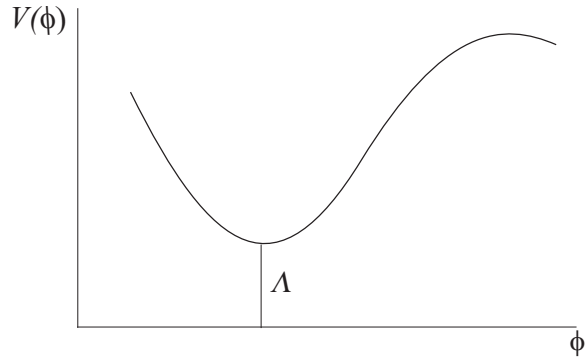}
	\caption{
		\footnotesize
		{Potential with a  local minimum and a positive cosmological constant }}
\end{figure}

The embedding of such a scenario in a  string theory cosmological framework is one of the key challenges today (see~\cite{Baumann:2014nda} for a review). 
String compactifications  are characterised by the appearance of a large number of moduli
 fields~\footnote{For reviews in string compactifications see~\cite{Douglas:2006es}-\cite{Blumenhagen:2006ci}.} and 
the question is  whether some of these moduli could be appropriate  inflaton candidates. 

In  string compactifications the ten-dimensional space-time is assumed to be a direct product
of the four-dimensional Minkowski  spacetime and a six-dimensional Calabi-Yau (CY)  manifold
characterised by a compact `radius' $R$. The classical supergravity equations, however, remain invariant under rescalings
of the  size $R$ of the compact  manifold.  Consequently, for any solution which determines four dimensional effective theory 
models such as the  Standard Model, we encounter a family of solutions by just changing the parameter    $R$.  In fact,
from the four-dimensional theory point of view,  $R$  corresponds to a massless scalar field. 

In general,   deformations of the compactification in four dimensions correspond to massless scalars which   do not acquire tree-level 
potential  and do not affect the four-dimensional action. Such scalars are the dilaton field, the K\"ahler and complex 
structure moduli, those corresponding to brane  deformations etc.  The appearance of such massless scalars in the effective theory
have singificant consequences. If  they couple gravitationally to matter fields, they could  mediate long range forces which 
have not been observed. They also affect the big-bang nucleosynthesis and have implications on the cosmological evolution.
 Therefore, to construct a realistic effective field theory model,  it is important to  generate a potential with 
 de Sitter vacuum  and assure a positive mass-squared for the various moduli at large enough volume to allow
 perturbative calculation. This is dubbed as the ``moduli stabilisation'' problem.   

Because there are a large number of  choices of 
CY manifolds and quantised fluxes which determine the properties of the effective 
field theory and, particularly, the scalar potential, 
string theory is characterised by a vast number of vacua, which constitute the so called
``string landscape''\footnote{For a recent update regarding  supporting evidence see also~\cite{Schellekens:2015zua}}.  However, 
whether there exist any de Sitter vacua amongst the plethora of possibilities is a long standing issue and the subject of an ongoing debate today. 
The  string computational tools used to  determine their existence are mainly the various moduli fields, the K\"ahler potential and the flux-induced superpotential 
in the effective field theory limit obtained after  compactification. Since
the tree-level potential for moduli fields vanishes identically, possible interesting non-zero contributions are based on $\alpha'$  and string loop corrections
as well as on non-perturbative effects. Notwithstanding the accumulated work and the variety of models that have appeared the last decades providing evidence
for the existence of the desired dS vacua,  most -if not all- proposed solutions   are based on assumptions and ingredients that might not be supported in a fully fledged string theory construction.  These doubts are corroborated
by the fact that a number of no-go theorems (see for example~\cite{Maldacena:2000mw,Covi:2008ea,Danielsson:2018ztv} ) preclude the existence 
of dS vacua, although at the classical level and under certain assumptions that may not be universally true  in string theory. 

 On the other hand, recently, some criteria on the (non)-existence of dS vacua in string theory have been proposed. The swampland conjecture suggested by the authors~\cite{Obied:2018sgi,Agrawal:2018own}  states that the  scalar
 potential $V$ of any effective field theory consistent with string theory
 satisfies  the bound
 \begin{equation}
 \frac{|\nabla V|}{V}\;\ge \;\frac{c}{M_{P}}\label{conj1}
 \end{equation} 
 where $c$ is a positive constant of order one.  This bound excludes  stable 
 or meta-stable de Sitter string vacua and, if true, it suggests that
 the latter  belong to the swampland. In a standard  inflationary scenario this essentially
 means that the slow-roll parameter $\epsilon$ is violated at order one.
 
 \noindent 
  Then, in the string framework the alternative interpretation
 of the present day value of the cosmological constant is through
 quintessence, provided the potential is positive, $V>0$, and 
  $|\nabla V|\sim V$~(for a review in quintessence see for example~\cite{Tsujikawa:2013fta} and in the context of strings~\cite{Cicoli:2018kdo}). 
  Counterexamples explaining how the bound ~(\ref{conj1}) 
  could be evaded have been proposed~(see for example~\cite{Andriot:2018wzk,Kehagias:2018uem,Denef:2018etk,Junghans:2018gdb}.
  However, the authors~\cite{Ooguri:2018wrx}  (see also~\cite{Garg:2018reu,Andriot:2018mav})
  proposed a refined version of the conjecture, according to
  which the potential must satisfy either the bound~(\ref{conj1}) or
 the  constraint associated with the minimum value of the mass spectrum:
 \begin{equation}
 {\rm min} (\nabla_i \nabla_j V )\;\le \; -\frac{c'}{M^2_{P}}\label{conj2}
\end{equation} 
Here, $c' $ is a positive   constant of order one.
 The  two bounds together, still forbid  de Sitter minima but allow the existence of maxima. 

In the ensuing years  after the experimental confirmation of the accelarating expansion of the universe, there has been a lot of
activity to construct effective theories of string origin with de Sitter vacua. However, the moduli space and the induced action
used to determine the four-dimensional vacua are not exactly known and the proposed models are based on assumptions regarding
the significance of the various quantum corrections.  Focusing on type IIB effective string theories in particular,  the usual procedure in 
obtaining vacua is based on flux compactifications with  $\alpha'$ and loop corrections playing a central r\^ole in moduli stabilisation. 
However, two of the most questioned ingredients  introduced  are the non-perturbative corrections in the superpotential and the uplifting term
(to ensure a dS vacuum) obtained from anti-$D3$  ($\overline{D3}$) branes. 

The focus of this talk is on this issue. That is, to examine  alternative solutions to the problems of moduli stabilisation and to
consolidate a de Sitter vacuum  based only on perturbative corrections.  The layout of the talk is organised as follows.
It will start with a short presentation of the string derived no-scale effective supergravity mainly focusing on the description of the moduli space,
the flux induced superpotential and the K\"ahler potential in the context of type IIB theory. Next, a short description of the r\^ole
of non perturbative effects on moduli stabilisation will be given and two familiar models proposed some 15 years ago will be briefly outlined. 
Afterwards, a new solution to the problem of moduli stabilisation based only on perturbative corrections in the presence of 7-branes will be 
presented in some detail.   The presentation will resume with  a brief exposition of the cosmological implications of this model.
More precisely it will be shown that, in the presence of a new Fayet-Iliopoulos term,  the slow-roll inflation  scenario can also be realised.

\section{The general type IIB set up}

In this talk we will discuss the issues of moduli stabilisation, de Sitter vacua and inflation in a type IIB/F-theory framework
~\footnote{For F-theory reviews see~\cite{Denef:2008wq, Heckman:2010bq,Weigand:2018rez}.}.
We will start with a presentation of the basic elements used to obtain the scalar potential of the effective $4d$ field theory, concentrating on the
bosonic spectrum and the moduli space.  The type IIB spectrum is obtained  by combining left and right moving sectors with Neveu-Schwarz (NS) 
and Ramond (R) boundary conditions. The bosonic spectrum  includes the graviton $g_{\mu\nu}$, a scalar   $\phi$ (the dilaton),  the two-index antisymmetric
 tensor $B_{\mu\nu}$-the so called  Kalb-Ramond (KB)  field, and the $p$-form potentials $C_p, p=0,2,4$. The axion and the dilaton fields are 
 combined to form the axion-dilaton 
 \begin{equation}
 S = C_0+i e^{-\phi} \label{axdi}
 \end{equation}

\noindent 
 In addition, various other geometric moduli
 corresponding to the deformations of the metric $g_{\mu\nu}$   emerge which are related to the shape, size and other properties of the   
 compactification manifold.
In particlular, these are classified as follows: 
\begin{enumerate}
	\item K\"ahler moduli $T^a$ corresponding to deformations of the K\"ahler form $J= i g_{i\bar j} dz^i \wedge d\bar z^{\bar j}\to T^a \omega_a$, 
	where $\omega_a,\,a=1,2,\dots, h^{1,1}$ is a basis for harmonic (1,1)-forms.
	
	\item Complex Structure moduli $z^b,\; b=1,2,\dots, h^{2,1}$  being harmonic (2,1)-forms, corresponding to deformations associated with a (3,0)-form, denoted with the greek letter $\Omega$. Clearly, the latter depends on the 
	complex structure moduli $\Omega=\Omega(z_a)$ 
	
	\item Scalar fields $b^a, c^a, \vartheta^a$ arising from the expansion of the 
	 $B_2, C_2, C_4$ potentials in the appropriate basis of harmonic forms.
	
	\item Moduli associated with   $D7$ brane deformations,  $D3$ positions etc. 
	
\end{enumerate}

In order to describe the four dimensional  effective supergravity limit of 
type IIB string, we need to derive the superpotential and the K\"ahler 
potential. 

\noindent 
In type IIB string theory the superpotential is generated by 3-form fluxes which are 
constructed from the KB-field and the $C_p$ potentials. The latter 
give rise to the field strengths
\begin{equation}
F_p:= d\, C_{p-1},\; H_3:= d\, B_2~,
\end{equation}
From these, the following combination is assumed
\begin{equation}
G_3:= F_3 -S\, H_3~,
\end{equation}
where $S$ is given in (\ref{axdi}). Using the holomorphic (3,0)-$\Omega$ form,
the flux induced superpotential is 
\begin{equation}
{\cal W}_0 =\int G_3 \wedge \Omega(z_a)~, \label{W0}
\end{equation}
while $G_3$ should be primitive $(2,1)$ form to allow ${\cal N}=1$ supersymmetry.  
The superpotential (\ref{W0}) depends on the compex structure moduli
$z_a$ through its $\Omega$ dependence but it does not involve the K\"ahler 
structure moduli. To ensure the flatness of the superpotential, we impose the supersymmetric conditions, setting zero the covariant derivatives 
with respect to the ${z_a}$  and $S$ fields:
\be  
{\cal D}_{z_a} {\cal W}_0=0,\;  {\cal D}_{S} {\cal W}_0=0~.
\label{Wflat}
\ee 
The solutions of these equations  stabilise the complex structure moduli and the axion dilaton field.
Both are assumed to obtain masses of the order of the string scale, while ${\cal W}_0$ becomes a constant.
However, because  the K\"ahler moduli do not participate in $ {\cal W}_0$,  
at this stage, they remain completely undetermined. 

\noindent 
Next we proceed with the second significant ingredient of the theory, namely the K\"ahler potential.
Ignoring for the moment any quantum corrections, the classical K\"ahler potential ${\cal K}$  is given by
the formula
\begin{eqnarray}
{\cal K}& =& -\sum_{i = 1}^3 \ln(-i({ T_i-\bar{T_i}}))\nonumber \\
&& - \ln(-i({ S-\bar{S}}))-\ln(i\int{\Omega}\wedge {\bar{\Omega}})~\cdot\label{K0}
\end{eqnarray}
In (\ref{K0}) we have assumed the contribution of three K\"ahler moduli fields, $T_i,\, i=1,2,3$
and $S$ is the axion-dilaton defined in~(\ref{axdi}). Moreover,   ${\Omega}={\Omega}(z_a)$ , is the holomorphic (3,0)-form~\cite{Candelas:1989js}, already 
introduced previously,  whilst $ {\bar{\Omega}}$ stands for its corresponding antiholomorphic one.

\noindent 
From (\ref{K0}), we can compute the scalar potential, using the standard supergravity formula
\begin{eqnarray}
{ V}&=&e^{\cal K}(\sum_{I,J} {\cal D}_I{\cal W}_0{\cal K}^{I\bar J}{\cal D}_{\bar J}{\cal W}_0-3 |{\cal W}_0|^2)\label{VfromK}
\end{eqnarray}
The sum is under all moduli fields, ($z_a, T_i, S$), while 
${\cal K}_{I\bar J}= \partial_I  \partial_{\bar J}  {\cal K}$ and ${\cal K}^{I\bar J}$ its inverse.  
Due to no-scale structure, at the classical level, this is identically zero and this can be easily seen by 
 splitting the sum into that of the K\"ahler moduli and the rest moduli fields: 
\begin{eqnarray}
{ V}&=&e^{\cal K} \sum_{ I, J =z_a,S{\neq T_i}}
D_I {\cal W}_0 {\cal K}^{I\bar J} D_{\bar J}{\cal W}_0\; \;({  D_I {\cal W}_0 =0},\,{\rm flatness} ) \nonumber \\
&&+e^{\cal K}\left(\sum_{ I, J = T_i}
{\cal{K}}_0^{I \overline{J}} \partial_I {\cal W}_0
\partial_{\overline{J}}{\cal W} _0 - 3 |{\cal W}_0|^2\right) \;\;({ =0}, \;{\rm no\,scale})\nonumber\\
&=&0\nonumber
\end{eqnarray}
The first line is zero due to supersymmetric conditions~(\ref{Wflat}) which fix the scalars  $z_a, S$ while the second line 
vanishes identically in no-scale models. Hence, at this level a scalar potential with zero vacuum energry is 
obtained and the K\"ahler moduli remain unfixed.

\section{Non Perturbative Corrections}

 Up to this point we have seen that it is  not possible to stabilise all 
 moduli and construct a potential with dS minimum at the classical level.  
 To circumvent this problem, most of the suggested solutions  rely on 
   non-perturbative  corrections to the superpotential . (Recall that
    perturative corrections in  ${ \cal W}_0$ are not allowed because of  non-renormalisation theorems.)
  For the sake of completeness of the presentation,
  two representative  solutions  will be briefly sketched, namely, ${\cal  A}$) the KKLT model and  ${\cal  B}$) the large volume scenario (LVS).

  \vspace{1cm} 
    
  \noindent
 Case ${\cal  A}$:\, In the simplest version of the model~\cite{Kachru:2003aw}, one assumes corrections of the form
\be 
{\cal W} = {\cal W}_0+A e^{ -\lambda T}
\ee
A possible  origin of these corrections come from a stack of $N$ 7-branes, wrapping 4-cycles, and it is associated
with ${\cal N}=1$ supersymmetric $SU(N)$ symmetry. In this case  gaugino condensation can take place and the constant $\lambda$ 
is given by $\lambda =2\pi/{ N}$, for $SU(N)$.
This way, the supersymmetric condition ${\cal D}_{T}{\cal W} =0$ stabilises the { $T$}-modulus, 
 however, this solution  requires { unnatural} fine-tuning of the parameters ${\cal W}_0,\, A$ and ${\lambda }$.
Moreover, the so obtained vacuum has an { AdS} (supersymmetric) minimum
\be
 {V_{AdS}} \propto  - 3 |{\cal W}|^2 e^{\cal K}\, { <}\, 0~,
\ee 
 in obvious conflict with the cosmological observations which imply $\Lambda>0$.  
 
 \noindent
  A solution to this problem  is obtained by   {\bf uplifting} the vacuum 
 to a ${\bf dS}$ minimum with the inclusion of a positive  contribution $V_{\overline{D3}}$ coming from ${\overline{D3}}$ branes, so 
 that the total contribution equals the cosmological constant 
\be 
V_{\overline{D3}}-|V_{AdS}| \approx 10^{-120} M_P^4 \label{VLambda}
\ee 
The 
$ \overline{D3}$'s source of  positive energy comes from 
$ {\overline{D3}}\, $ tension and the 3-form fluxes are introduced to cancel  the Tadpole.
Indeed, for $N_3 (\bar N_3)$  number
of $D3 (\overline{D3})$ branes the charge $Q_3=N_3-\bar N_3$ is determined by the global tadpole condition 
\ba 
\frac{\chi{(X)}}{24}&=& Q_3+ \frac{1}{2\kappa^2_{10}{\cal T}_3}\int H_3\wedge F_3~,
\ea 
where, in an F-theory framework,  ${\chi{(X)}}$ is the Euler characteristic of the CY fourfold $X=CY_4$,
and ${\cal T}_3$  is the tension of the $D3$-brane. Now, according to  Klebanov-Strassler's description~\cite{Klebanov:2000hb}, 
charge conservation is imposed by assuming appropriate $M$ and $-K$ units of charge $\int_A F_3= 4\pi^2 M,\; \int_B H_3= -4\pi^2 K$,
where the integrals are over the $A$-cycle and its dual $B$-cycle respectively. In addition, the fluxes  generate 
a superpotential for the complex structure moduli~\cite{Giddings:2001y}.
The   positive energy of the $\overline{D3}$ depends on the warp factor $e^{ A}$:
\[  V_{\overline{D3}} = 2{\mu}e^{4A},\; \; ds_{10}^2 = e^{2 A}dx_4^2+e^{-2A}dy_6^2~.
\]
However, as pointed out in the above references (see~\cite{Giddings:2001y,Kachru:2002gs}),
 the supersymmetry preserved by the  $\overline{D3}$
branes is incompatible with the global supersymmetry preserved by the imaginary self-dual
3-form flux of the background geometry and as a result,  this configuration is  unstable.

Another way to  parameterise the uplifting of an AdS vacuum of the potential to a de Sitter one, is by using  nilpotent chiral multiplets. 
In this version of the KKLT construction   a  goldstino multiplet $S_{NL}=s+\sqrt{2} \theta G+\theta^2 F_S$ 
 satisfying the constraint $S_{NL}^2=0$ associated with non-linearly realised supersymmetry of the Volkov-Akulov type is 
introduced~\footnote{for recent studies of nilpotnet goldstino see~\cite{Komargodski:2009rz,Antoniadis:2010hs,Antoniadis:2014oya}.}. 
In this framework, the superpotential and the K\"ahler potential are~\cite{Kallosh:2006dv,Ferrara:2014kva}
\ba 
{\cal W}&=& {\cal W}_0 +A\, e^{-\lambda T}+\mu^2 S_{NL}\label{WS}\\
{\cal K}&=& -3\,\ln(T-\bar T)+S_{NL}\bar S_{NL}
\ea 
where, as above, $T$ is the K\"ahler modulus associated with the volume and $S_{NL}$ the nilpotent goldstino superfield $S_{NL}^2=0$.
Computing the scalar potential, 
one finds~\footnote{In
the warping case we have instead~${\cal K}= -3\,\ln(T+\bar T+S\bar S)$ and $V_{up}= \frac{\mu^4}{(T+\bar T)^2}$.}
\be 
V_{\rm new}= V_{AdS}+ V_{up}=V_{AdS}+\frac{\mu^4}{(T+\bar T)^3}\label{Vnew}
\ee 
Therefore, the existence of a nilpotent goldstino implies an uplifting term exactly as the one obtained 
by the $\overline{D3}$ brane. The important difference is that now this is manifestly supersymmetric. 
However, given the doubts whether these features can be part of a realistic string scenario and the amount of fine-tuning required, 
the importance  of these effects on the determination of a true de Sitter vacuum remains distinctly nebulous. 

\vspace{1cm} 
  
\noindent
 Case ${\cal  B}$:  
The    Large Volume  Scenario (LVS)~\cite{Balasubramanian:2005zx} can be thought as a generalisation of the above 
constructions, aiming to  improve some of the aforementioned deficiencies. This proposal is also based on the assumption of
non-perturbative corrections to the superpotential but it  is  realised with an exponentially large volume. 
In the simplest  version  the volume  is given  by ${\cal V}= \tau_b^{3/2}-\tau_s^{3/2}$ where   $\tau_b, \tau_s$, 
( for a generalisation see~\cite{Cicoli:2008gp}) 
are two distinct K\"ahler moduli.  The superpotential  and K\"ahler potential are given by
 \ba
  \mathcal{W}_{LVS} &&= \mathcal{W}_0 + A\, e^{- \lambda\tau_s}~,\label{LVS}\\
  \mathcal{K}_{LVS} &&=  - 2 \textrm{ln} (\tau_b^{\frac{3}{2}} - \tau_s^{\frac{3}{2}} + \xi)~.
 \ea
The new ingredient in the K\"ahler potential is a constant $\xi$  which arised  due to  $\alpha^\prime$ corrections, and it is proportional 
to the Euler number~\cite{Candelas:1990rm}
\be 
\xi=-\frac{\zeta(3)}{4(2\pi)^3}\chi\label{xi}
\ee  
Both,  gaugino condensation and  the $\alpha^\prime$ correction { $\xi$} are required to stabilise the K\"ahler moduli ${\tau_b} ,{\tau_s}$.
However, as in   KKLT scenario (case ${\cal A}$),  a mechanism is required to uplift the potential to a { dS}  minimum. This can be done by  
 introducing gauge field fluxes on $D7$-branes which induce a D-term potential in the effective four dimensional model~\cite{Burgess:2003ic}.

\section{Perturbative moduli stabilisation}

In the rest of this talk, an alternative scenario of K\"ahler moduli stabilisation  will be presented
which does not rely on any uncontrolable non-perturbative terms in the superpotential. Earlier computations~\cite{Berg:2004ek,Berg:2007wt} focused on
 one-loop corrections in the string coupling and contributions to order ${\alpha'}^3$~\cite{Becker:2002nn}  breaking the no-scale  invariance
 of the K\"ahler potential. However, it was realised that these are not capable of stabilising the volume and  the K\"ahler moduli $T_i$ in general,
 unless a $T_i$-dependent superpotential is assumed.   Instead, the proposed mechanism can be realised in a type IIB/F-theory
 framework and relies on the presence  of logarithmic corrections  induced by $D7$-branes~\cite{Antoniadis:1998ax} in  the 4d effective action.
At the end of this talk it will be shown how the incorporation of a new Fayet-Iliopoulos (FI)-term~\cite{Fayet:1974jb}
will allow for a slow-roll inflation.

To be more specific, a geometric configuration of three intersecting 7-branes with  fluxes  will be considered with the corresponding
K\"ahler moduli fields  denoted as $T_a= \tau_a+ i b_a$. We begin with the description of the basic ingredients. 
It is known that  when ${\cal O}({\alpha'}^3)$ corrections are taken into account, the definition of 
the four-dimensional dilaton $\phi_4$, in terms of the ten-dimensional one,  is given by
\be 
e^{-2\phi_4}= e^{-2\phi}\left({\cal V}+\xi\right)\label{4dil}
\ee 
where ${\cal V}$ is the CY volume and $\xi$ the constant~(\ref{xi}) which is proportional to the Euler number. 
To express the volume in terms of the K\"ahler moduli we introduce the following notation:\\
  We denote with $v^a$  the 2-cycle volume modulus transverse to the $D7_a$ brane
 and, assuming three intersecting branes,  the total volume is given by 
\be 
{\cal V} = \frac 16 \kappa_{abc} v^av^bv^c~,\label{6v}
\ee  
where $\kappa_{abc}$ are intersection numbers.  The world volume associated with the $D7$ brane is defined by
\be 
\tau_a =\frac 16 \kappa_{abc}v^bv^c ~,\label{4v}
\ee 
Taking into account the ${\alpha'}^3$ corrections in type IIB, the K\"ahler potential is given by the formula 
\be 
{\cal K}= -2\ln (e^{-2\phi} ({\cal V}+\xi) )- \ln\left(\int \Omega\wedge\bar{\Omega}\right)+{\rm constant}
\ee 
In order to convert (\ref{4dil}) in the Einstein frame, we define
\be 
\hat v_a =  v_a e^{-\phi/2}=\frac{\tau_a}{g_s^{1/2}}, \;\; \hat{\xi}= \xi e^{3\phi/2}= {\xi}{g_s^{3/2}}~.
\label{tauxi}
\ee 
Then, in terms of (\ref{tauxi}), expression (\ref{4dil}) is written
\[e^{-2\phi_4}= e^{-\frac 12\phi}\left(\hat{\cal V}+\hat\xi\right)~,
\]
where $ \hat{\cal V}=\frac 16 \kappa_{abc} \hat v_a\hat v_b\hat v_c$.
Then the  K\"ahler potential takes the form:
\begin{equation}
{\cal K}= -\ln ({ S-\bar S}) -2\ln (\hat{ \cal V}+\hat\xi )
		- \ln\left(\int \Omega\wedge\bar{\Omega}\right)+{\rm constant}~,	\label{KahlerX}
\end{equation}
where use has been made of the relation $S-\bar S \propto e^{-\phi }$ derived from (\ref{axdi}).

The specific form (\ref{KahlerX}) of the K\"ahler potential in type IIB can be  confirmed by a T-duality 
transformation~\cite{Antoniadis:2018hqy} from that of type IIA.

\subsection{Logarithmic loop corrections and D-terms from $D7$-branes}

$D7$ and $D3$ branes  are essential elements in configurations aiming to describe viable effective field
theories derived from type IIB flux compactifications   and its F-theory geometric analogue. 
Their deformations are  important ingredients of the moduli space and could be useful to ensure a successful inflationary
scenario and  a stable de Sitter vacuum. 
In such configurations, branes which span different dimensions of the compact space intersect each other 
and, depending on the details of the geometry and fluxes, may be associated with anomalous $U(1)$ symmetries. 
Then,  the resulting chiral spectrum of the four dimensional
effective theory  induces Fayet-Iliopoulos  D-terms in the effective potential. Based on these observations,
in the following, we will examine the r\^ole of intersecting D7 brane configurations in moduli stabilisation, de Sitter vacua and 
slow roll inflation. 

We start with the observation  that in the large volume limit, a nonvanishing $\xi$ correction
 gives rise to localised graviton kinetic terms in the  Calabi-Yau compact manifold at the points where the Euler number 
 is concentrated.  As discussed in \cite{Antoniadis:2002tr}, this implies a localised Einstein action and 
 there is an emission of closed strings from the associated graviton vertices into the various  $D$ branes
of the assumed geometric configuration, giving rise to local tadpoles. In the case of $D7$ branes,  closed strings propagating in the 
the two-dimensional transverse space induce an infrared divergence which exhibits a logarithmic dependence  
in the regime of large transverse volume of codimension two~\cite{Antoniadis:1998ax}. Due to the infrared divergence, 
one could also expect that it is the dominant correction at that order in the string loop expansion. For the simplest case 
of a single $D7$ brane, denoting the size of the transverse dimensions
with $u$,  the corresponding loop  correction takes the form  
\begin{equation}
\delta =\gamma \ln(u)~,\label{D7transcor}
\end{equation}
where $\gamma$  is a model dependent  parameter. These corrections, together with the ${\alpha}'$ corrections discussed
earlier, appear in  Einstein kinetic terms in ${\cal S}$:
\[ { \cal S} =-\frac{1}{2\kappa^2_{4} } \, \int d^{4} x \sqrt{-g} \left(  e^{-2{\phi }}({\cal V} +
 { \xi})+{ \delta}\right)
 { {\cal R}}+\cdots
\] 
The K\"ahler potential takes the form~\cite{Antoniadis:2018hqy}
\begin{equation}
{\cal K}= -\ln ({ S-\bar S}) -2\ln (\hat{ \cal V}+\hat\xi+\hat{\delta} )
		- \ln\left(\int \Omega\wedge\bar{\Omega}\right)+{\rm constant}~,	\label{KahlerXD}
\end{equation}
   where $\hat{\delta}= \delta g_s^{1/2}$. 
  One can notice that both $\xi$ and $\delta$ corrections break the no-scale form of the K\"ahler potential. 
 \begin{figure}[h!]
	\centering
	\includegraphics[width=0.45\columnwidth]{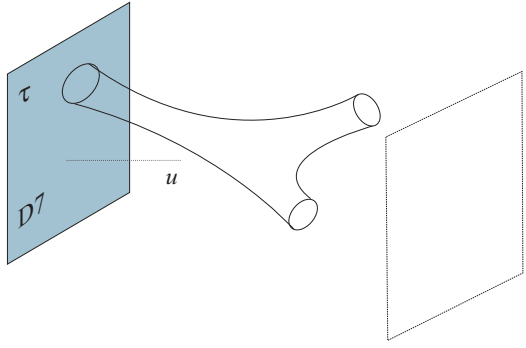}
	\caption{
		\footnotesize
		{ Sting loops from exchange of closed strings between graviton vertices and $D7$-branes when $\xi\ne 0$.  }}
\end{figure}

\noindent
Next, we describe the second important effect of the $D7$ branes which  are  associated with anomalous $U(1)$ symmetries. 
As already mentioned, there is an  induced D-term which has the generic form dictated by the effective ${\cal N}=1$ 
supergravity \cite{Burgess:2003ic, Jockers:2004yj, Achucarro:2006zf, Haack:2006cy}:
\ba
V_D = \frac{g_{D7}^2}{2} \left( iQ\partial_{T^a}{\cal K}(T^a)+\sum_JQ_J\mid\langle \phi^J\rangle \mid^2 \right)^2\label{D7Kahl}
\ea 
The gauge coupling is fixed by the kinetic function: $\frac{1}{g^2_{D7}}={\rm Im}(T^a)$ and  $\phi^J$ are scalar components
of superfields whose charges $Q_J$ are subject to anomaly cancellation conditions, 
which  are automatically satisfied in a consistent string background~\cite{Antoniadis:2004pp,Antoniadis:2005nu}.
Although in general the VEVs of the scalar fields may be non-zero, for our
present purposes we can ignore the matter fields and write (\ref{D7Kahl})
as follows
\ba
V_D=-\frac{d_a}{{2\rm Im}(T^a)} \left(\partial_{T^a}{\cal K}(T^a)\right)^2,\label{VDterm}
\ea 
in which $d_a = Q^2$.

\begin{figure}[h!]
	\centering
	\includegraphics[width=0.4\columnwidth]{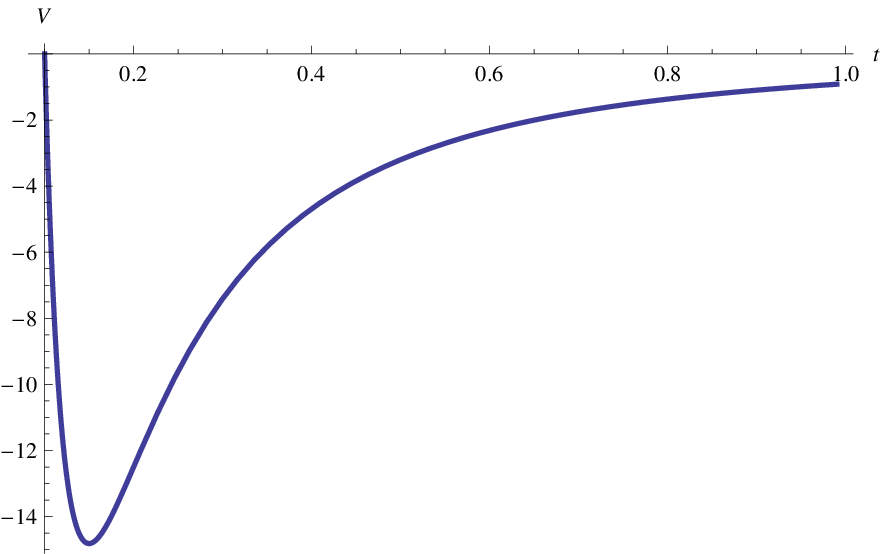}\;
	\includegraphics[width=0.45\columnwidth]{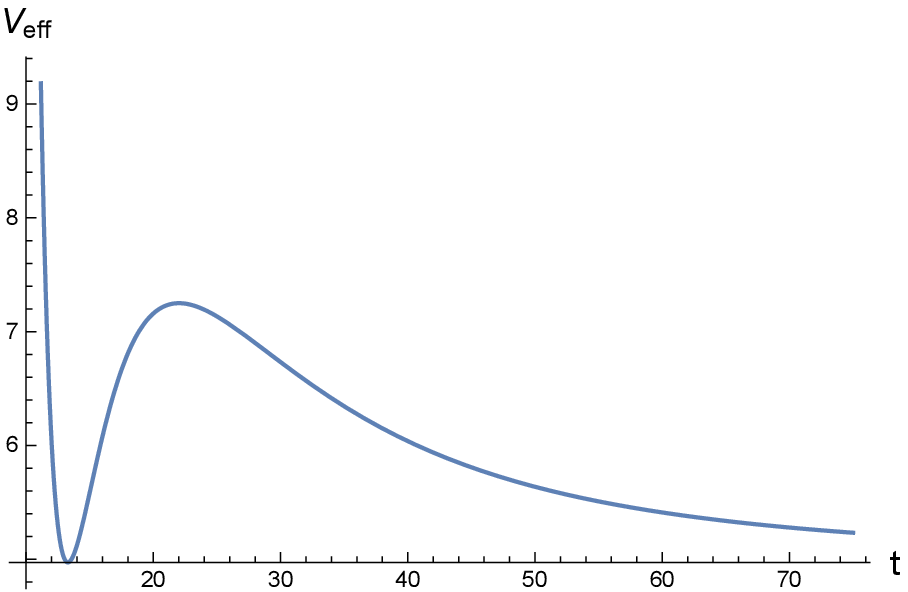}
	\caption{Anticipated shapes of the scalar potential including moduli-dependent perturbative corrections
	plotted vs   the modulus $t$.\,	Left: a typical potential with AdS minimum. Right: Imposing 
	the condition $\lim_{t\to \infty}V(t) \to 0$  and the requirement of de Sitter minimum, imply also a maximum at some finite $t$.}
	\label{Vgen}
	\end{figure}

Before we start a detailed analysis of the model, it is important to set out the basic features required for
an acceptable dS vacuum.  We have seen that at the classical level the effective potential vanishes due to
no-scale properties and flatness conditions. The inclusion of perturbative moduli-dependent 
quantum corrections in the K\"ahler potential should induce contributions to the scalar potential, $ V(t)$
with $t=(\tau, u)$. The validity of perturbation theory implies that such corrections should vanish for $t\to \infty$
and therefore $\lim_{t\to \infty}V(t) \to 0$. If the zero at infinity is reached from negative values, then, 
 for reasonable potentials, this implies an AdS minimum which is not acceptable. 
 Thus, the vanishing of the potetnial at infinity should be approached 
from positive values. Again, for non-contrived structures of the potentials~\footnote{For more involved cases see for example~\cite{Kallosh:2006dv}.}, this implies that
there should be somewhere a maximum before a dS minimum is  formed. This is plotted in figure ~\ref{Vgen}.

 \subsection{ $D7$ Branes and Moduli Stabilisation}

Having determined the form of the perturbative quantum corrections arising  in the presence of ${ D7}$ branes,
we can investigate their r\^ole in the effective theory.
It can be shown~\cite{Antoniadis:2018hqy}  that the stabilisation of the K\"ahler moduli requires at least three magnetised 7 branes intersecting each other. 
 Therefore, we consider here directly the effects of  three $D7$ branes.
 We introduce three K\"ahler moduli $T_1, T_2, T_3$ and express the internal 6d volume in terms of their imaginary parts $\tau_k=\frac{1}{2i}(T_k-\bar T_k)$.
 	We first recall that the volume is given by (\ref{6v}). Given also the relation (\ref{4v}), we can write 
 	\be 
 	{\cal V}= (\tau_1\tau_2\tau_3)^{\frac 12}~\cdot\label{V6t}
 	\ee
 	In terms of (\ref{V6t}),  the K\"ahler potential is written as follows
 	\ba
 	{\cal K}&=& -2\ln\left( (\tau_1\tau_2\tau_3)^{\frac 12}+\xi+\sum_{k=1}^3 \gamma_k \ln(\tau_k)\right)~,\label{Kah123}
 	\ea 
 	where $\gamma_k$ are model dependent coefficients of order one. For simplicity we take 
 	$\gamma_1=\gamma_2=\gamma_3\equiv \gamma_{\tau} $,  and absorb the $\alpha'$ corrections, $\xi$, into the logarithmic  term
 	by a new parameter $ \mu = e^{\frac{\xi}{2\gamma_\tau}}$ so that 
	 \be 
	 \xi + 2 \gamma_{\tau } \ln({\cal V}) = 2 \gamma_{\tau } \ln(\mu{\cal V})~.
	  \ee
	 The F-term potential is 
 	\be 
 	V_F= 3\gamma_{\tau}  W_0^2   \frac{\
 		2( \gamma_{\tau}  +2{\cal V}) +(4\gamma_{\tau} -{\cal V})\ln(\mu {\cal V})}
 	{({\cal V}+2 \gamma_{\tau}   \ln(\mu {\cal V}))^2 \left(6 \gamma_{\tau} ^2+{\cal V}^2+8\gamma_{\tau} 
 	 {\cal V}+\gamma_{\tau}  (4 \gamma_{\tau}  -{\cal V}) \ln(\mu {\cal V})\right)}~,\label{VF}
 	\ee
 	and in the large volume limit with small logarithmic corrections,  can be  approximated by 
 	\be 
 	V_F\approx 3 W_0^2\,\gamma_{\tau} \, \frac{\ln(\mu {\cal V})-4}{{\cal V}^3}~\cdot \label{potV}
 	\ee 
 D-term 	contributions to the effective potential  in the presence of  $D7$  fluxed branes take the form
 	\ba 
 	V_D&=&\sum_{a=1}^3\frac{d_a}{\tau_a}\left(\frac{ \partial{\cal K}}{\partial \tau_a}\right)^2
 	\approx \sum_{a=1}^3\frac{d_a}{\tau_a^3}
	~,
 	\label{VD}
 	\ea 
where the approximation holds 	in the large volume limit.
	
 	\noindent
Hence, when both, F- and D-term contributions are taken into account, in the large volume expansion,  the effective potential is  
 as	follows
 	\[ V_{\rm eff}\;\approx\; 3 W_0^2\,\gamma_{\tau} \,
 	 \frac{\ln(\mu {\cal V})-4}{{\cal V}^3}\,+ \,\frac{d_i}{\tau_i^3}\,+\,\frac{d_j}{\tau_j^3}\,+\,\frac{d_k(\tau_i\tau_j)^3}{{\cal V}^6}, \]
  for $ i\ne j\ne k\ne i$. 
 \noindent	
 	  There are  three independent variables  $\tau_{1,2,3}$ and the product of them is related to the 6d-volume.
 	We replace one of the $\tau_i$'s with the total volume ${\cal V} $ and  minimise the potential with respect to
  ${\cal V} $ and the	two remaining K\"ahler moduli. 
 	Two minimisation conditions determine the ratios between the moduli,  $\left(\frac{\tau_i}{\tau_j}\right)^3=\frac{d_i}{d_j} $
 	and the third one the total volume.
 	Expressed in terms of the stabilised total volume ${\cal V}$, the conditions for the two $\tau_i$ can be written as
 	\[\tau_i^3 = \left(\frac{d_i^2}{d_kd_j}\right)^{\frac 13}{\cal V}^2~\cdot  \]
 	The minimisation condition for the total volume reads:
 		\begin{eqnarray}
 	 	 \frac{13}{3} -\ln{\cal V}\;=\;  \frac 23 \frac{ d}{\gamma}\, {\cal V}~.\label{Vdef}
 	 	\end{eqnarray}
 	Inserting the three conditions in $V_{\rm eff}$ we obtain the simple form
 	\be 
 	V_{\rm eff}\approx  \gamma \, \frac{\ln(\mu {\cal V})-4}{{\cal V}^3}+\frac{d}{{\cal V}^2}~,
 	\ee 
 	with 
 	\ba 
 	d=3(d_1d_2d_3)^{\frac 13}~\label{defd}\;;\;
 	\gamma=3W_0^2\gamma_{\tau}~\cdot \label{defgamma} 
 	\ea 

\noindent 	
 The above scalar potential possesses a dS minimum in some region of the parameter space~\cite{Antoniadis:2018hqy}. 
To show this, we introduce the convenient  definition:
\begin{eqnarray}
{ w} &= &\frac{13}{3} -\ln{\cal V}~.\label{Vdefw}
\end{eqnarray}
Then, vanishing of the derivative of the potential with respect to ${\cal V}$ (i.e. condition \ref{Vdef}), takes the form:
\ba
 w e^{{ w}} \;{=}\; { z}~.\label{wze}
\ea
\begin{figure}[t!]
	\centering
	\includegraphics[width=0.4\columnwidth]{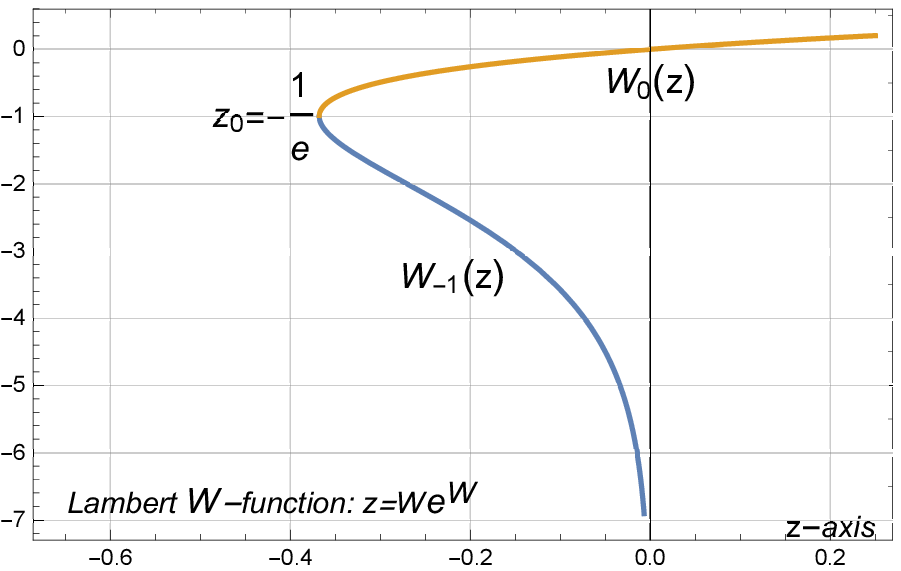}\;\;\;
	\includegraphics[width=0.4\columnwidth]{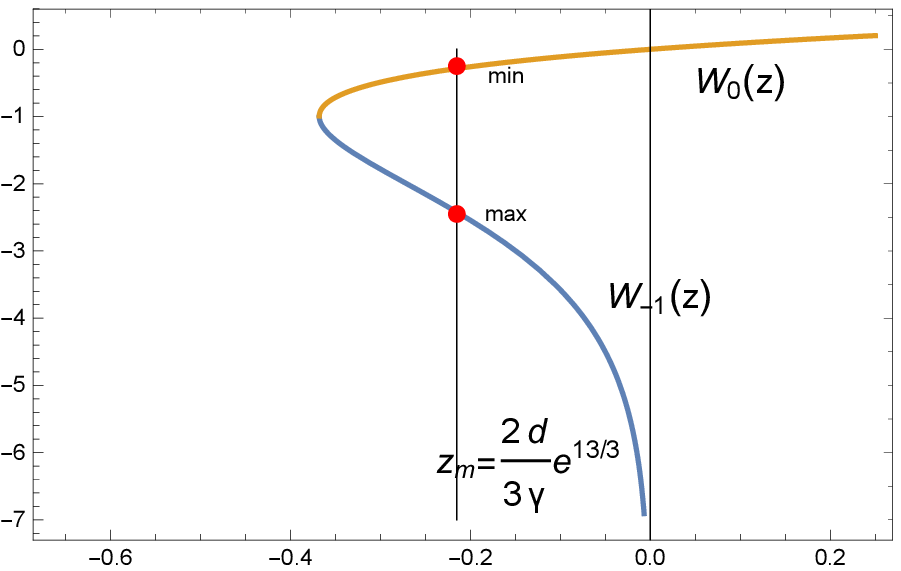}
\caption{Left: The two branches of the Lambert $W$-function,  $W_0$ and $ W_{-1}$,  respectively. Right:
The two values of the world volume ${\cal V}$ for fixed $d/gamma$ ratio, corresponding to the minimum and the maximum  of the  scalar potential.}	
\label{LambertW01}
	\end{figure}
where, in the present case the variable $z$ is related to the ratio of the two constants,
 \be 
 {z} = \frac 23 \frac{ d}{\gamma}e^{\frac{13}3}~. \label{zdgamma}
 \ee 
 Inversion  of  (\ref{wze}) determines ${ w}$ and  consequently ${\cal V }$  through (\ref{Vdefw}).
The final solution given by  
 multivalued Lambert $W$-function~\cite{LambertW} 
\be 
w\Rightarrow { W}({ z})~.\label{wtoW}
\ee
The two branches of the Lambert function  ${ W_0}({ z}) $ and  ${ W_{-1}}({ z}) $
are shown in figure~\ref{LambertW01}. 
Real values of  ${ W_0}, { W_{-1}}$  are obtained  for  $ { z}\ge { z_0}=-{ e}^{-1}$.

	\noindent 
Furthermore, the values of the parameters $d$ and $\gamma$, whose ratio determines $z$, must be such that 
the scalar potential $V=V({\cal V})$ has a minimum and a maximum w.r.t. the volume ${\cal V}$, or equivalently w.r.t.
to the function $W(z)$ as can be inferred from their relation, see~(\ref{Vdefw}) and (\ref{wtoW}).  
Thus, we are interested for solutions such that $z$ (determined from the ratio of parameters $d$ and $\gamma$, see Eq~(\ref{zdgamma})), 
falls in the range where, simultaneoulsy two solutions for the volume can be obtained. In the right hand side of figure \ref{LambertW01} the
 vertical line represents any value of ${ z_m} = \frac{2{ d}}{3{\gamma}}e^{\frac{13}3}$
between 
\be 
-{ e}^{-1}\le  \frac{2{ d}}{3{\gamma}}e^{\frac{13}3}\le 0~,\label{doublesol}
\ee 
 where  $V_{min}$ and $V_{max}$ can coexist. The maximum of the scalar potential 
corresponds to the intersection of the $z_m$ line with the branch $W_{-1}(z_m)$ whilst the minimum
is determined by the intersection of $z_m$ with $W_{0}(z_m)$.

The  requirement for  de Sitter vacua puts additional restrictions.   To implement this
constraint, we first compute the minimum value of the scalar potential. The volume at the 
minimum is 
\[ {\cal  V}_{0} = e^{\frac{13}3-{W_0}(\frac{2}{3}\frac{d}{\gamma}e^{\frac{13}3})}\]
We substitute ${\cal  V}_{0}$ in ${ V_{\rm eff}=V_F+V_{D}}$    and  require $V_{\rm eff}$ to   be positive: 
\[{V_{\rm eff}^{min}} = \frac{\gamma}{{\cal V}_0^3}+ \frac{ d}{{\cal V}_0^2}>0
\]
This constraint puts an upped bound on the ratio $\frac{d}{\gamma}$. Combined  with the lower 
bound  on $z$ in~(\ref{doublesol}), we obtain the range
\be
- 7.242 <10^3 \frac{d}{\gamma} < -6.738~\cdot 
 \label{dgammabound}
 \ee

\section{Slow roll inflation} 

As is well known, slow-roll inflation occurs by a scalar field $\phi$, dubbed the inflaton,
 which starts from the top of the potential $V(\phi)$ rolling down slowly compared to the expansion of the Universe. 
 The equation of motion is 
 \be 
 \ddot{\phi} +3 H \dot{\phi} +V'(\phi)=0\,,\label{EoM}
 \ee 
 where the derivative $V'$ is w.r.t $\phi$. The expansion of the universe $H=\frac{\dot a}{a}$ is involved in the friction term and  
 its derivative implies  $\frac{\ddot a}{a}=\dot H+H^2= H^2(1-\varepsilon)$, where $\varepsilon= -\frac{\dot H}{H^2}$ is one of the
slow roll parameters.  In order to have accelarated expansion, $\ddot a>0$, this must be bounded in the region $0<\varepsilon <1$. 
Furthermore, to ensure slow rolling with almost constant velocity of  the inflaton along
the potential, a second condition should be imposed: $\ddot{\phi}\ll |H\dot{\phi}|$. This is associated 
with a second slow roll parameter $\eta =-\frac{\ddot{\phi}}{H\dot\phi}<1$. Under the above conditions 
the equation of motion (\ref{EoM}) can be approximated $3 H\dot{\phi}\approx -V'(\phi)$. 

\begin{figure}[h!]
	\centering
	\includegraphics[width=0.70\columnwidth]{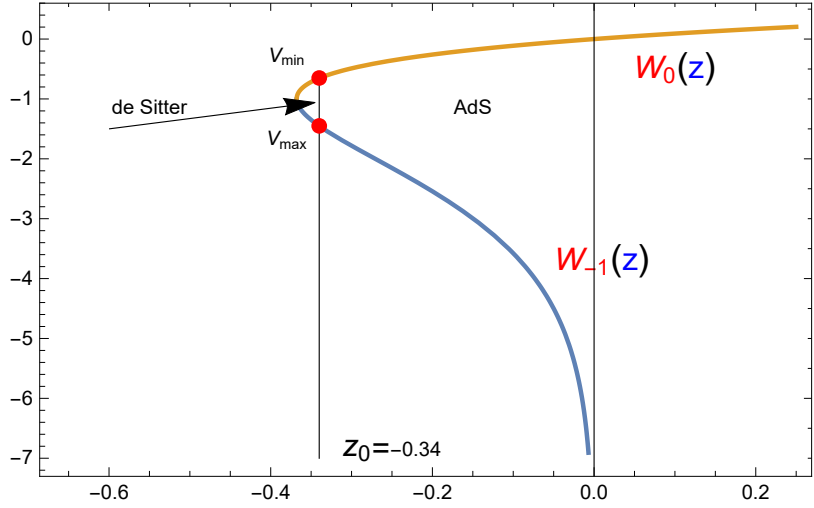}
	\caption{The allowed region for a dS minimum. The vertical line $z_0=-0.34$ defines the maximum `distance' 
	of the two extrema of the scalar potential $V_F+V_D$ consistent with a dS vacuum.}
	\label{dSAdS} 
	\end{figure}

To study  cosmological inflation in the present model, we are led to identify the inflaton field with the 
compactification volume modulus ${\cal V}$. To this end, in order to obtain  a canonically normalised kinetic term,
 we perform  a suitable transformation of the K\"ahler moduli fields. 
It turns out that the appropriate transformation is~\cite{Antoniadis:2018ngr}
\be 
t_i = \frac{1}{\sqrt{2}}\ln (\tau_i)~,
\ee
Next, we switch  to the normalised real scalar fields and
define a convenient basis  in terms of the normalised volume $t$ :
\ba
t &=&\frac{1}{\sqrt{3}}(t_1 + t_2 + t_3) = \frac{\sqrt{6}}{3} \ln ({\cal V})~,\label{ttoV}
\ea
and two  perpendicular directions
\ba 
u &=&  \frac{1}{\sqrt{2}} (t_1 - t_2)~,\label{ufi}\\
v &=&  \frac{1}{\sqrt{6}} (t_1 + t_2 - 2t_3)~\label{vfi}\cdot 
\ea 
In order to have a model consistent with an effective theory with positive cosmological constant,
we would like to accommodate the slow-roll inflation and at the same time to ensure a dS vacuum. 
However, although a dS minimum exists, the actual allowed region of the parameter space is too restrictive, see Fig.~\ref{dSAdS},
and additional requirements  for  slow-roll inflation such as the number of efoldings are hard to be met.
Therefore, in the context of the present construction, the implementation of  the inflationary scenario 
would not be possible unless  a suitable uplifting term is included in the scalar potential~\cite{Antoniadis:2018ngr}.  

\noindent 
A possible  way out of this impasse is a novel Fayet-Iliopoulos (FI)-term
proposed in ~\cite{Cribiori:2017laj}. This term  is gauge invariant at the Lagrangian level and 
can be written for a non-R $U(1)$ symmetry~\cite{Antoniadis:2018cpq}. With this in mind, we introduce a constant term  $V_{up}$ associated to a $U(1)$ of a $D3$-brane, so 
that the effective potential takes the form~\cite{Antoniadis:2018ngr}:  
 	\ba
 V_{\rm eff}   
 &=&\gamma \, \frac{1}{2} e^{-\frac{3\sqrt{6}}{2}t} (\sqrt{6}t - 8) + d\,e^{-\sqrt{6} t}  + V_{up}~,\label{VVup}
 	\ea
 with the modulus $t$ related to the total volume via~(\ref{ttoV}) and $\gamma$  defined in eq. (\ref{defgamma}).
 	
\begin{figure}[h!]
 		\centering
 		\includegraphics[width=0.80\columnwidth]{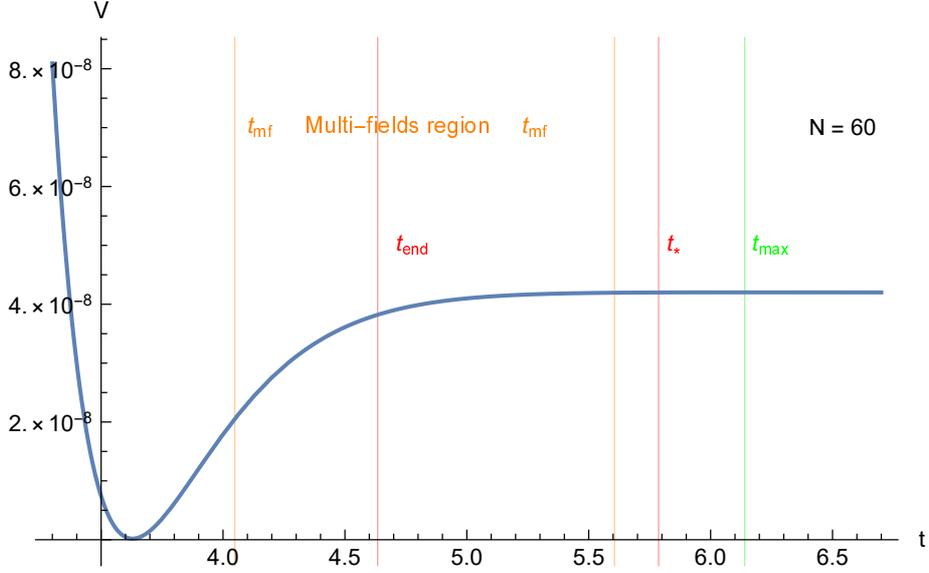}
 		\caption{
 			\footnotesize
 			{The scalar potential for   $n_s = 0.9605, N_* = 60$}. Inflation occurs in the region $(t_*, t_{end})$ which lies 
 			between the maximum and the minimum of the potential. The region on the right of this line is characterised by AdS minima.}
 			\label{N60}
 	\end{figure}

 Inflation should occur in an interval $t=(t_{end}, t_*)$ which lies  between the maximum  and the minimum  of the scalar potential. 
 The first end of the interval is the ending point of inflation, which corresponds  to the breaking of the slow-roll condition:
  	\be
  	t_{end} ={\rm max}\left[t |_{\frac{1}{2}\left(\frac{V'}{V}\right)^2 \simeq 1}, t |_{\frac{|V''|}{V} \simeq 1}\right], 
  	\ee
  	where the derivatives $V', V''$ are taken with respect to $t$. The second point is the one corresponding to the 
  	anticipated number of e-foldings
  	 $N_*\sim 50$ to 60, where $N_*$ is given by the formula
  	\be
  	 N_* = \int_{t_{end}}^{t_{*}} \frac{V}{V'}dt~\cdot
  	 \ee
  	In addition, at the same point $t_* (N_*)$, the spectral index should satisfy the values from observations
  	\be
  	n_{s} = 1-6\epsilon+2\eta=1 - 3\left.\left(\frac{V'}{V}\right)^2\right|_{t_*} + 2\left.\frac{V''}{V}\right|_{t_*}~\cdot\label{spectralindex}
  	\ee
   In Fig.~\ref{N60}, the case  for $N_* = 60$ efoldings is shown. As required,  the region $(t_*, t_{end})$ where infaltion takes place is stretched out 
    			between the maximum and the minimum of the potential. Because of the presence of two other scalar fields (\ref{ufi}, \ref{vfi}),
    			multi-field effects should be considered in the region where the inflaton field is no longer the lightest scalar (see~\cite{Antoniadis:2018ngr}
    			for a detailed analysis\footnote{See also recent work~\cite{Achucarro:2018vey} for multi-field infation.}).

\section{Conclusions}

The quest for the existence of de Sitter string vacua in string theory is a fascinating subject of intensive  recent reseach.
Contradictory conjectures with greater or lesser theoretical foundation regarding their existence abound, and the results
are far from being conclusive.  Moduli fields which are always present in Calabi-Yau compactifications, play an instrumental
r\^ole in the determination of  the effective theory's  vacuum. 

In this presentation an effective  model in a type IIB framework  is considered based on a geometric configuration with 
three intersecting stacks  of $D7$-branes. In  this set up, perturbative string loop  contributions  induce terms in the K\"ahler potential
which depend logarithmically on the volume moduli  associated with the directions transverse to the corresponding $D7$-branes. 
 In this framework, the K\"ahler moduli fields are stabilised and a de Sitter vacuum is accommodated naturally. 
 In the final part of this presentation, the implications on inflation are discussed.

  \section*{Acknowledgements}
   	This work was supported in part by the Swiss National Science Foundation, in part by Labex ``Institut Lagrange de Paris'' and in part 
   	by a CNRS PICS grant. 	G.K.L. would like to thank the Corfu school organisers for the kind invitation and 
   	the  LPTHE in Paris for  kind hospitality  where part of the work was completed.


\end{document}